\documentclass[notitlepage,pra,twocolumn,longbibliography,amsmath,amssymb]{revtex4-1}  

\usepackage{graphicx}
\usepackage{xcolor}                              
\definecolor{linkblue}{rgb}{0,0,0.65}            
\usepackage[colorlinks=true,            % use colored text links
linkcolor=linkblue,             % internal links (eq refs, etc.)
citecolor=linkblue,             % citation links
urlcolor=linkblue]{hyperref}    % external links
\pdfoutput=1

\makeatletter
\let\oldthebibliography\thebibliography
\renewcommand\thebibliography{\let\textbf\relax\oldthebibliography}
\makeatother

\begin{document}
	
\title{Wavefunction branches demand a definition!}

\date{\today}

\author{C.~Jess~Riedel}\email{jessriedel@gmail.com}\affiliation{Physics \& Informatics Laboratories, NTT Research,\ Inc., Sunnyvale, CA, USA}

\begin{abstract}
	Under unitary evolution, a typical macroscopic quantum system is thought to develop wavefunction branches: 
	a time-dependent decomposition into orthogonal components that (1) form a tree structure forward in time, (2) are approximate eigenstates of quasiclassical macroscopic observables, and (3) exhibit effective collapse of feasibly measurable observables. 
	If they could be defined precisely, wavefunction branches would extend the theory of decoherence beyond the system-environment paradigm and could supplant anthropocentric measurement in the quantum axioms.
	Furthermore, when such branches have bounded entanglement 
	and can be effectively identified numerically, 
	sampling them would allow asymptotically efficient classical simulation of quantum systems.
	I consider a promising recent approach to formalizing branches on the lattice by Taylor \& McCulloch [\emph{Quantum} 9, 1670 (2025), arXiv:2308.04494], and compare it to prior work from Weingarten [\emph{Found.\ Phys.}\ 52, 45 (2022), arXiv:2105.04545]. 
	Both proposals are based on quantum complexity and argue that, once created, branches persist for long times due to the generic linear growth of state complexity.
	Taylor \& McCulloch characterize branches by a large difference in the unitary complexity necessary to interfere vs.\ distinguish them.
	Weingarten takes branches as the components of the decomposition that minimizes a weighted sum of expected squared complexity and the Shannon entropy of squared norms.
	I discuss strengths and weaknesses of these approaches, and identify tractable open questions.
\end{abstract}

\maketitle

Consider a huge many-body system and divide it into some macroscopic variables S (system) and the remaining variables E (environment). Suppose S is initially localized in phase space, i.e., the joint wavefunction $\psi(0)$ of S+E is an approximate eigenstate of a complete set $\hat{S} = \{\hat{X}_a^{\mathrm{S}},\hat{P}_a^{\mathrm{S}}\}$ of canonically conjugate observables on S, $\hat{S} \psi(0) \approx s_0 \psi(0)$. What happens under unitary evolution by macroscopic chaotic dynamics? The wavefunction will evolve into a superposition of eigenstates of $\hat{S}$ with macroscopically different eigenvalues \cite{vonneumann2010Proof, goldstein2010LongTime,buniy2021Macroscopic} after just several multiples of the characteristic Lyapunov time \cite{chirikov1988quantum, zurek1998decoherence}, which may be only seconds or minutes. If S is approximately Markovian because E is large and weakly coupled, decoherence \cite{zurek2003decoherence,joos2003decoherence,schlosshauer2008decoherence,zurek2025decoherence} quite generally \cite{hernandez2023decoherence,hernandez2024classical} induces a time-dependent decomposition into orthogonal components, 
\begin{align}
	\psi(t) = \sum_{i\in B(t)} \psi_i(t), \qquad \hat{S} \psi_i(t) \approx s_i(t) \psi_i(t),
\end{align}
that for a long time proliferate and \textbf{branch} in this specific sense: a component at a later time $t_2$ has non-negligible overlap with only one forward-evolved component from an earlier time $t_1<t_2$, ensuring they form a (rooted) tree graph \cite{footnoteBrevity}. 
Although multiple branches $\psi_i(t)$ may share the same $\hat{S}$ eigenvalue at a given time, all branches are orthogonal because they have distinct states of E. 
%This implies \textbf{non-interference} of operators on S: the expectation value is the Born-weighted mean of the expectation values on the branches.
This implies \textbf{effective collapse} for operators on S: the expectation value is the Born-weighted mean of the expectation values on the branches. 
The branches are \textbf{quasiclassical} because the approximate eigenvalues of $\hat{S}$ along any trajectory drawn from the tree approximately follow the classical equations of motion with noise \cite{hernandez2023decoherence,hernandez2024classical}. If the system contained a laboratory experiment, the different measurement outcomes would correspond to disjoint subsets of the branches.

%%%%%%%%%%%%%%%%%%%%%%%%%%%%%%%%%%%%%%%
%   Begin Branching figure
%%%%%%%%%%%%%%%%%%%%%%%%%%%%%%%%%%%%%%%

\begin{figure}[t]
	\includegraphics[width=\linewidth]{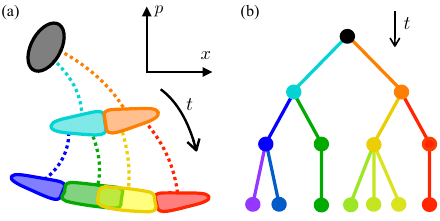}
	\caption{(a) A schematic Wigner function of diverging chaotic trajectories of S in phase space, continuously decohered by E. Distinct trajectories may arrive at overlapping states of S (green and yellow), but they correspond to orthogonal branches because they are associated with different conditional states of E, i.e., E contains a record of the past of S. (b) Branches of S+E at different times form a rooted tree graph defined by their time-evolved overlap with earlier and later branches.}
	\label{fig:branching}
\end{figure}

%%%%%%%%%%%%%%%%%%%%%%%%%%%%%%%%%%%%%%%
%   End Branching figure
%%%%%%%%%%%%%%%%%%%%%%%%%%%%%%%%%%%%%%%

This narrative assumes a fixed division of the universe into intuitive subsystems: S vs.\ E, and parts thereof. But the most obvious subsystems, macroscopic bound objects, are transient, forming at one time and later breaking apart. Bound objects are a special case \cite{gell-mann2007quasiclassical} of \textbf{hydrodynamic} variables, i.e., local averages of exactly and approximately locally conserved quantities like momentum and species number, and perhaps best generalized to slow local operators \cite{kim2015Slowest}. The consistent histories formalism provides a flexible language for describing the decoherence of these more general variables without an eternal S-E split \cite{halliwell2003decoherence, gell-mann1993classical, gell-mann2007quasiclassical}.

But the transience problem persists: the appropriate choice of variables and degree of historical coarse-graining changes in time as conditions change, e.g., as the universe cools, and as winter submits to spring.  This ambiguity becomes acute in isolated thermalizing systems, since finer histories interfere earlier. The histories \emph{framework} can be readily adapted to shifting variables in a branch-dependent way \cite{gell-mann1990Quantum,gell-mann2013adaptive}, but there's no systematic \emph{prescription} for picking variables and degree of coarse-graining; human expertise is required on a case-by-case basis.  Indeed, the appropriate variables of a many-body system are emergent, generally not being easily deducible from the Hamiltonian \cite{anderson1972more,laughlin2000theory}.

Thus one can \emph{sketch} a unified story to describe the quasiclassical behavior of everyday macroscopic systems but -- even putting computational issues aside -- we don't have a \emph{procedure} for filling in the blanks.  We can seek comfort in arguments that this is unobjectionable or even inevitable \cite{griffiths2014New,wallace2003Everett}, but deep down \cite{dowker1996consistent,okon2014Measurements} you know it's unsatisfying!

\section{The Microscopic Perspective}

So suppose I didn't describe any of that in terms of subsystems, preferred variables, or ``macroscopic''. Instead, imagine you're just handed a complete microscopic description of the joint wavefunction of all the atoms. In particular, the representation treats each atom on equal footing, and no one has told you separately which one is adrift in the air and which one is part of a coffee cup (or even that there \emph{is} a coffee cup). Per the account above, we still expect that this wavefunction is well-approximated as a time-dependent sum of orthogonal components, each of which is quasiclassical in some sense.   But without being told what the preferred variables are, how would you systematically identify them, even approximately?  What Python script would you write whose input was the wavefunction and whose output was the set of branches?

This question has appeared in various guises going back to Everett, especially as the problem of identifying a preferred set of consistent histories; for reviews, see the introductory sections of  Refs.~\cite{kent2014solution,riedel2016objective,carroll2020quantum,weingarten2021macroscopic,zampeli2022contrary,ollivier2022emergence,strasberg2023everything}. The framing here in terms of branches is a choice that emphasizes (a) the effective equivalence of sets of histories \cite{gell-mann1990Alternative,gell-mann1994Equivalent} that correspond to the same set of branches and (b) the possibility that branches are best identified by examining the entangled state at a given time rather than the trajectory or Hamiltonian.

Is this problem even well-posed? Why expect a  \emph{unique} answer rather than a different decomposition for every set of variables that catches our fancy?

A seemingly humble desideratum about any macroscopic fact is that it ought to be ``objective'' in the sense that many different observers could deduce it through independent local measurements \cite{zurek2009quantum, zurek2025decoherence}; if it's not recorded in multiple places, like our brains, how else could we even be talking about it?  This intuition can be formalized without any reference to observers per se, but rather just in terms of the state's correlations across disjoint spatial regions. Under relatively mild assumptions that the regions aren't too delicate or extended, one can show that a wavefunction can be \emph{uniquely} decomposed into simultaneous eigenstates of \emph{all} such objective observables \cite{riedel2017classical}. Unfortunately, the most obvious ways of formalizing ``delicate'' require choosing a length scale (a recurring theme; see below).  And importantly, this decomposition probably doesn't form a tree structure in time, nor can its components recover the apparent bounded entanglement of the macroscopic world. Nevertheless, it suggests that spatial locality provides sufficient foundation to build a quasi-unique branch decomposition from the instantaneous wavefunction without knowing which subsystems or variables are preferred.

\section{The Promise of Branches}

Why bother with branches if the full wavefunction is sufficient for an outside observer to make predictions?  Consider:
\begin{enumerate}
	\item 
	Decoherence theory has proven insightful and practically useful; it's worth generalizing to cases where the preferred variables vary or are unknown.
	\item 
	Bounded-order correlation functions could be classically computed by \textbf{sampling branches} \cite{riedel2017classical, taylor2025nonInterfering}, potentially reducing or even (if they have area-law entanglement \cite{footnoteAreaLaw}) eliminating the exponential growth rate of simulation cost.
	This would naturally extend the efficient classical simulability of decohering variables \cite{wiseman2014Quantum,hernandez2023decoherence} beyond the open-system paradigm.
	\item
	The branches $\psi_i$ and probabilities $|\psi_i|^2$ would give a unified description of \emph{what happens} in closed systems like the universe, recovering the traditional collapse postulate in the special case of lab measurements.
\end{enumerate}

Think of the stakes! A principle that picks out a wavefunction's distinct outcomes -- of cosmic inflation, of last week's storm, and of the photodiode in your lab -- would convert an initial state of the universe into a time-dependent menu of macroscopically interpretable possibilities. A comprehensive definition would hardly eliminate all the mystery of the infamous measurement problem, but it would remove a fundamental ambiguity.  No longer would collapse be evasively defined using the eigenbasis of whatever observable was measured, implicitly appealing to human intuition to carve the world into measured subsystem and measuring device.

Solving this would mean finding universal criteria that identify a time-dependent decomposition of a many-body wavefunction into orthogonal components that
\begin{itemize}
	\item form a tree structure in time, until at least thermalization;
	\item are approximate eigenstates of quasiclassical macroscopic observables like hydrodynamic variables; 
	\item exhibit effective collapse of feasibly measurable observables; and 
	\item have bounded (perhaps area-law) entanglement.
\end{itemize}  
\emph{This} is the problem of defining wavefunction branches, and it demands to be solved!

%%%%%%%%%%%%%%%%%%%%%%%%%%%%%%%%%%%%%%%
%   Begin Complexity figure
%%%%%%%%%%%%%%%%%%%%%%%%%%%%%%%%%%%%%%%

\begin{figure}[t]
	\includegraphics[width=\linewidth]{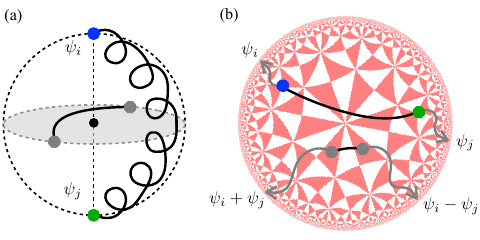}
	\caption{(a) The Bloch sphere for the span of $\psi_i$ and $\psi_j$. They are Taylor-McCulloch branches when low-complexity operations can distinguish $\psi_i$ and $\psi_j$ but only high-complexity operations can interfere them, i.e., distinguish the states $\psi_i + e^{i\theta}\psi_j$ lying in the equatorial disk. Any operation that swaps $\psi_i$ and $\psi_j$ (curly black path) is necessarily complex, but states on opposite sides of the disk can be swapped with low complexity (smooth black path). (b) Quantum complexity induces a metric on Hilbert space with strong negative curvature. As two states evolve in time, they effectively head in opposite directions in the sense that the least complex path connecting them will approximately retrace both trajectories. A large initial difference between the swap complexities of two pairs of states is preserved.}
	\label{fig:branching}
\end{figure}

%%%%%%%%%%%%%%%%%%%%%%%%%%%%%%%%%%%%%%%
%   End Complexity figure
%%%%%%%%%%%%%%%%%%%%%%%%%%%%%%%%%%%%%%%

\section{A New Hope}

Recently, Weingarten \cite{weingarten2021macroscopic} and Taylor \& McCulloch \cite{taylor2025Wavefunction} have made tentative but fascinating proposals for defining  branches from first principles.  Both proposals are fundamentally based on \textbf{quantum unitary complexity} $\mathcal{C}(U)$ \cite{yao1993quantum, nielsen2006quantum,baiguera2025quantum}, a quantification of how complicated a unitary operator $U$ is, given roughly by the length of its simplest decomposition into elementary operations. The complexity of the least-complex unitary mapping between  two states $\psi$ and $\phi$ then defines their \textbf{quantum state complexity} $\mathcal{C}(\psi,\phi)$, a distance in Hilbert space. Preferred variables are completely eschewed; the only scaffolding is spatial locality, which enters through the choice of elementary operations, e.g., nearest-neighbor. 

The second law of quantum complexity \cite{susskind2016typicalstate,brown2017quantum,brown2018second,baiguera2025quantum} -- that state complexity rises nearly linearly and maximally under generic local Hamiltonian evolution -- is invoked to infer how branches behave over time. This potentially connects branching (effective collapse) directly to physical irreversibility, addressing a deficiency in Ref.~\cite{riedel2017classical}. While there is some freedom in picking the set of elementary operations (e.g., nearest-neighbors vs.\ 2-local), desirable symmetries like translation invariance and local-unitary invariance dramatically cut down the options. Less obvious desiderata in the continuum limit may narrow things further \cite{camargo2019path}.  Many different choices of elementary operations at very short distances become equivalent on larger scales, yielding a small number of equivalence classes \cite{brown2023universality,brown2024polynomial}.

Working carefully in lattice field theory, Weingarten defines branches as the decomposition $\psi = \sum_i \psi_i$ that minimizes a weighted sum of the expected squared complexity of the branches and the Shannon entropy of their squared norms, 
\begin{align}
	Q(\{\psi_i\}) := \sum_i |\psi_i|^2 [\mathcal{C}(\psi_i,\Omega)^2- b \ln |\psi_i|^2],
\end{align}
where $\Omega$ is the vacuum state and $b$ is a free parameter. If quantum state complexity (relative to the vacuum)  correctly measures how branches are less complicated than the overall state, then Weingarten’s decomposition -- minimizing $Q(\{\psi_i\})$ -- is a natural guess; it optimizes for the smallest amount of branch uncertainty that achieves the lowest expected branch complexity. The additive form and choice of Shannon entropy are both driven by the reasonable desideratum that uncorrelated spatial regions branch independently.

Weingarten gives a careful preliminary analysis of how such branches will behave and how they can be made Lorentz covariant.  He proves detailed upper and lower bounds on the complexity of some states, including a field-theoretic version of the GHZ state which branches when the GHZ subsystems extend beyond a length scale associated with $b$.  Indeed, Weingarten shows that, at least in a few examples, his definition is self-consistent in the continuum limit if $b$ scales like it has units of volume.  He finds that correlations on each branch probably cannot extend much beyond the length scale associated with $b$, which would enforce area-law entanglement but at the expense of potential tension with long-baseline interferometry. He gives an extensive, although necessarily heuristic, argument that the second law of complexity ensures that a low-entropy initial state branches only forward in time, assuaging ``unbranching'' concerns \cite{footnoteWeingarten}.

Taylor \& McCulloch, working on a fixed non-relativistic lattice, instead propose that branches be characterized by a decomposition that is \emph{hard to interfere} but relatively \emph{easy to distinguish} with local operations. More precisely, they say $\psi = \sum_i \alpha_i \psi_i$  is a decomposition into ``good'' branches, up to small error $\epsilon$, if  $\mathcal{C}_{\mathrm{I}}^{i,j}-\mathcal{C}_{\mathrm{D}}^{i,j} \gg 1$ for all $i\neq j$, where $\mathcal{C}_{\mathrm{D}}^{i,j}$ and $\mathcal{C}_{\mathrm{I}}^{i,j}$ are the complexities of the least complex unitaries $U_{\mathrm{D}}$ and $U_{\mathrm{I}}$  that achieve
\begin{align}\label{eq:tmDcond}
	\frac{|\langle \psi_i | U_{\mathrm{D}} |\psi_i\rangle-\langle \psi_j | U_{\mathrm{D}} |\psi_j\rangle|}{2}  & \ge 1-\epsilon,\\
	\label{eq:tmIcond}
	\frac{|\langle \psi_i | U_{\,\mathrm{I}} |\psi_j\rangle|+|\langle \psi_j | U_{\,\mathrm{I}} |\psi_i\rangle|}{2} & \ge \epsilon.		
\end{align}
Unitaries satisfying 
%\eqref{eq:tmDcond} 
the first equation can control arbitrary processes conditional on $\psi_i$ vs.\ $\psi_j$.  Similarly, by Aaronson et al.'s operational connection between swappability and interferability of quantum states \cite{aaronson2020hardness}, $\psi_i$ and $\psi_j$ cannot be interfered without effective access to unitaries satisfying 
%\eqref{eq:tmIcond}.
the second equation.

We can also interpret 
%\eqref{eq:tmDcond} and \eqref{eq:tmIcond} 
these equations 
through the lens of error correction: within a codespace spanned by the orthonormal basis $\{\psi_i\}$ of branches, the \emph{classical} information associated with the basis is recoverable \cite{knill2000theory} (up to error $\epsilon$) from unitary noise of complexity up to $\mathcal{C}_{\mathrm{I}} := \min_{i,j\neq i}\mathcal{C}_{\mathrm{I}}^{i,j}$, but the conjugate phase information is maximally \emph{not} recoverable from unitary noise of complexity $\mathcal{C}_{\mathrm{D}} := \max_{i,j\neq i}\mathcal{C}_{\mathrm{D}}^{i,j}$. This is particularly interesting in light of recent progress on codespace complexity in approximate quantum error correction, e.g., Ref.~\cite{yi2024complexity}.

Taylor \& McCulloch confirm that their definition behaves sensibly on GHZ states, product states, random-circuit states, and some quantum codes. They go on to present heuristic arguments that their branches are stable (i.e., remain good branches) on exponentially long timescales due to the second law of complexity and in particular the known behavior of precursor complexity. They also note that their branches will tend to have less entanglement than the overall state, as expected, due to the asymmetric effect entanglement has on  $\mathcal{C}_{\mathrm{I}}$ vs.~$\mathcal{C}_{\mathrm{D}}$.  Perhaps most interestingly, they appeal to an eigenstate thermalization hypothesis on low-complexity operators to argue that approximate eigenstates of conserved quantities are likely to branch, hinting at the possibility of recovering and generalizing the decoherence of quasiclassical hydrodynamics.

Taylor \& McCulloch's criteria are arguably more physically motivated than Weingarten's, but disconcertingly their decomposition isn't even approximately unique for a given $\psi$. As they show, multiple incompatible ``good" decompositions can coexist without being coarse-grainings of a common fine-grained decomposition. Interpreting each branch decomposition $\psi=\sum_i \alpha_i \psi_i$ as a code for protecting the classical information associated \cite{knill2000theory} with the commuting algebra $\mathcal{B}_{\{\psi_i\}} := \bigoplus_i \mathbb{C}|\psi_i\rangle\langle\psi_i |$, one wonders whether these can be unified in a single code for hybrid quantum-classical information associated with some non-commuting algebra $\bar{\mathcal{B}}$ that contains all the $\mathcal{B}_{\{\psi_i\}}$ as subsets.

In contrast, Weingarten defines a unique and exact branch decomposition at each time step by construction, consistent with the goal of establishing an unambiguous basis of macroscopic reality. Thus, up to a single free parameter, he offers an exact answer to our original question.  But is it the correct answer? I worry that minimization is a way to \emph{force} a precise decomposition, and that a sharp threshold may be incongruous with most models of decoherence, where off-diagonal elements decay exponentially and where effectively irreversible entanglement builds up rapidly but smoothly in the environment.  Because the free parameter $b$ controls the trade-off between the number and mean complexity of Weingarten's branches, one might suppose that varying $b$ just controls this approximation by coarse-graining the branches. But this does not seem to be what happens, and ostensibly conflicts with $b$ carrying a length scale. 

Because Taylor \& McCulloch work only on a fixed lattice, their proposal naively requires choosing a preferred length scale (the lattice spacing) if applied to quantum fields. Unlike Weingarten, they do not address the continuum limit. However, because they always consider complexities between components of the overall state, rather than relative to a vacuum state, this leaves open the optimistic possibility that UV divergences could cancel or be absorbed into the ``branchiness'' quantity, $\mathcal{C}_{\mathrm{I}}-\mathcal{C}_{\mathrm{D}}$, without picking a scale.

Taylor \& McCulloch's proposal involves a choice of error $\epsilon$ because they work with traditional circuit complexity rather than Nielsen's geometricized version. Given that we expect branches to be emergent and smooth, it's reasonable that this should not have sharp values, but little is known about how the branch structure behaves as the error is varied. I also wonder how varying the error (or, better, dispensing with it altogether by geometricizing their proposal) interacts with the branchiness quantity  $\mathcal{C}_{\mathrm{I}}-\mathcal{C}_{\mathrm{D}}$, e.g., is it possible to increase both the branchiness and the error to get roughly the same branch structure, or are these more like orthogonal axes?

Taylor \& McCulloch's proposal is also missing a relativistic generalization. Thus, this proposal requires significantly more development before it could be considered to be a candidate for a fundamental definition.

A well-known issue with quantum complexity is that it's generally infeasible to compute for generic states. This is an obstacle to developing numerical simulation techniques based on classically sampling over either Weingarten or Taylor-McCulloch branches, although heuristic and approximate approaches informed by such definitions can still work \cite{taylor2025nonInterfering}.  And because quantum complexity quantifies hypothetical trajectories between quantum states, complexity-based branch decompositions naturally blur the line between decompositions based on the instantaneous wavefunction $\psi$ and those based on the Hamiltonian or the state's full trajectory.

Are there states for which the Taylor \& McCulloch and Weingarten decompositions are dramatically different? (Only Weingarten's criteria depend on branch norms.) Can either type unbranch or otherwise violate the tree structure, contrary to heuristic arguments?  Does either clearly reproduce the decoherence of hydrodynamic variables? Or are these proposals heading in the wrong direction altogether?

\section{Onward}

A good definition of branches would allow us to precisely ask and answer fascinating questions we can currently only gesture at: how fast do branches form? In what sense are they discrete vs.\ continuous? What quantum computing resources are needed to simulate (and potentially reverse) a purportedly irreversible process? When do branches stop forming or recohere, and does this occur before, after, or concurrently with thermalization? After all, regardless of whether the Hilbert space of the universe is formally infinite dimensional, it is \emph{effectively} finite-dimensional in any finite region with finite energy, and so can only fit a finite number of orthogonal branches. 

The theory of decoherence has been celebrated for illuminating how the appearance of the classical world arises in our fundamentally quantum universe. But is the project \emph{complete}?  No! We are compelled to understand, and the path forward beckons.

\section*{Acknowledgments} 

I thank Adam Brown, Steven Hsu, Dan Ranard, Jordan Taylor, Akram Touil, and Don Weingarten for thoughtful comments.

\bibliography{references}

\begin{thebibliography}{54}
\bibitem [{foo({\natexlab{a}})}]{footnoteBrevity}%
  \BibitemOpen
  \href@noop {} {}\bibinfo {howpublished} {For brevity I will elide many
  important details throughout, like discrete vs.\ continuous branches,
  different flavors of consistent histories, rare re-interference, small
  non-orthogonalities, and the grounding or modification of the Born rule}
  \BibitemShut {NoStop}%
\end{thebibliography}
	
\end{document}